\documentclass[twocolumn,prb,showpacs,preprintnumbers,amssymb,superscriptaddress]{revtex4}
\usepackage{amsmath}
\usepackage{graphicx}
\usepackage{dcolumn}

\usepackage{bm}% bold math

\begin{document}
\title{Effects of thermal fluctuations on the magnetic behavior 
of mesoscopic superconductors}
\author{A. D. Hern\'{a}ndez}
\email{alexande@cab.cnea.gov.ar} 
\altaffiliation{Present addresss: The Abdus Salam International Centre for Theoretical Physics,
Strada Costiera 11, (34014) Trieste, Italy}
\affiliation{Centro At\'{o}mico Bariloche, 8400 San Carlos de
Bariloche, R\'{\i}o Negro, Argentina}
\author{B. J. Baelus}
\email{ben.baelus@ua.ac.be} 
\altaffiliation{Present addresss: Institute of Materials Science,
University of Tsukuba, Tsukuba 305-8573, Japan}
\affiliation{Departement Fysica, Universiteit Antwerpen (Campus Drie Eiken), \\
Universiteitsplein 1, B-2610 Antwerpen, Belgium}
\author{D. Dom\'{\i}nguez}
\email{domingd@cab.cnea.gov.ar}
\affiliation{Centro At\'{o}mico
Bariloche, 8400 San Carlos de Bariloche, R\'{\i}o Negro, Argentina}
\author{F. M. Peeters}
\email{francois.peeters@ua.ac.be}
\affiliation{Departement Fysica, Universiteit Antwerpen (Campus Drie Eiken), \\
Universiteitsplein 1, B-2610 Antwerpen, Belgium}

\date{\today}

\begin{abstract}
We study the influence of thermal fluctuations on the magnetic
behavior of square mesoscopic superconductors. The strength of
thermal fluctuations are parameterized using the Ginzburg number,
which is small ($G_i \approx 10^{-10}$) in low-$T_c$
superconductors and large in high-$T_c$ superconductors ($G_i
\approx 10^{-4}$). For low-$T_c$ mesoscopic superconductors we
found that the meta-stable states due to the surface barrier have
a large half-life time, which leads to the hysteresis in the
magnetization curves as observed experimentally. A very different
behavior appears for high-$T_c$ mesoscopic superconductors where
thermally activated vortex entrance/exit through surface barriers
is frequent. This leads to a reduction of the magnetization and a
non-integer average 
number of flux quanta penetrating the superconductor.
The magnetic field dependence of the probability for the
occurrence of the different vortex states and the fluctuations
 in the number of vortices are studied.

\end{abstract}

\pacs{74.78.Na, 74.40.+k, 74.20.De} \maketitle

\section{Introduction}

In the last decade, the study of vortex matter in finite
mesoscopic superconductors attracted a lot of attention, both
experimentally\cite{LN373m, LPRB60v, LN408m, LPRL86m,GN390, GN396,
GN407, Lindelof,kanda_baelus1,kanda_baelus2} 
and theoretically.\cite{PRB57v,PRL81v,PRL83v,PRB59,PPRL84,PC369p,PRB65,HD1,HD2}
Most of the experimental studies
on mesoscopic superconductors deal with conventional low-$T_c$
superconductors. Resistivity measurements are widely used to
obtain information on the superconducting/normal 
transition.~\cite{LN373m, LPRB60v, LN408m, LPRL86m} 
With ballistic Hall magnetometry it is possible to measure deep inside
the superconducting phase diagram and to obtain information on the
magnetization of the different superconducting states.\cite{GN390, GN396, GN407, Lindelof} 
Recently, A. Kanda et al. developed the multiple-small-tunnel-junction method
to distinguish between multivortex and giant vortex states in
mesoscopic superconductors. \cite{kanda_baelus1,kanda_baelus2} 
Theoretically, studies of the Ginzburg-Landau free energy
\cite{PRB57v,PRL81v,PRL83v,PRB59,PPRL84,PC369p,PRB65} 
and the time dependent Ginzburg-Landau  equations\cite{HD1,HD2}
have shown how the magnetic and dynamics properties  
depend on the sample sizes and geometry.
In particular the mesoscopic samples can develop Abrikosov multivortex 
states \cite{PRL81v} and depending on the size of the sample 
it is possible to observe first or second order 
transitions.\cite{PRB57v} 
One interesting
characteristic of the magnetic properties of mesoscopic superconductors is the
behavior of the dc magnetization curves. In mesoscopic superconductors 
there is a reinforcement of the surface 
barrier for entrance and exit of vortices.\cite{PRL83v,PRB59,HD1}
The surface barriers allow for the existence of meta-stable states of constant 
vorticity as a function of magnetic field and lead to a
magnetization curve with discontinuous  jumps, which have
been observed experimentally.\cite{GN390}  
All these results are for conventional low-$T_c$ superconductors,
where the effects of thermal fluctuations are generally small.

A very different situation occurs in macroscopic high-$T_c$
superconductors where the effects of thermal fluctuations are
important. In particular, fluctuations affect the dynamics of
vortex entrance through the surface barriers.
\cite{burlachkov91_exp,burlachkov94_teo,schilling,civale,niderost}
Recently, there has been interest in the study of the magnetic
properties of micron-sized high-$T_c$ superconductors
\cite{wang,kogan-clem}. Single fluxoid transitions and two-state
telegraph noise was measured in thin film rings at temperatures
close to $T_c$.\cite{kogan-clem} The fluxoid transitions were
explained by thermally activated penetration of vortices.

The magnetic and dynamic behavior of mesoscopic superconductors is
greatly influenced by surface barriers and geometry
effects.\cite{PRL83v,PRB59,PRB65,HD1} As a
consequence, thermal fluctuations in mesoscopic high-$T_c$
superconductors could lead to interesting phenomena due to thermally
activated vortex entrance/exit through surface barriers.

In the present paper we study the effects due to thermal
fluctuations on the vortex entry in square mesoscopic
superconductors. We solve numerically the time-dependent
Ginzburg-Landau equations taking into account demagnetizing
effects and thermal noise fluctuations. The strength of thermal
noise fluctuations is parameterized by the Ginzburg number. For
conventional low-$T_{c}$ superconductors the Ginzburg number is
small, while for high-$T_{c}$ superconductors this parameter is
large. We compare results obtained from the minimization of the
mean field free energy, corresponding to the equilibrium states in
the absence of thermal fluctuations, with the results from the
time dependent Ginzburg Landau (TDGL) dynamics with small Ginzburg
number (corresponding to conventional low-$T_c$ superconductors)
and the results for TDGL dynamics with large Ginzburg number
(corresponding to high-$T_c$ superconductors). The magnetization,
the number of vortices, the free energy, etc are calculated in
each case. We find very good agreement between the results of free
energy minimization and the results of TDGL with small Ginzburg
number. In the case of a large Ginzburg number the TDGL approach 
gives different and new results for the magnetization and the
vortex dynamics.

The paper is organized as follows. In Section II we give the two
theoretical formalisms used in the present paper. First we describe how
we solve the TDGL equations, obtaining the magnetic fields via the
Biot-Savart law and including thermal fluctuations. Secondly, we
describe the solution of the stationary Ginzburg-Landau equations, when
the magnetic field is solved via Ampere's law. In Section III the
results of the TDGL theory are compared with the ones of the stationary
GL theory. In a first step, we neglect the thermal fluctuations and
compare the results both at zero and finite temperature. In a second
step, we include thermal fluctuations in our calculations and compare
the results with small and large Ginzburg number. Finally, in
Section IV we present our conclusions.

\section{Theoretical formalism}

\subsection{Time-dependent Ginzburg-Landau theory}

We solve the time-dependent Ginzburg-Landau equations, in the
gauge where the electrostatic potential is zero, and taking into
account thermal fluctuations (see Ref. \onlinecite{enomoto,hohenberg}), i.e.
\begin{eqnarray} \Gamma^{-1}_{\Psi}\frac{\partial\Psi}{\partial t}&=& -
\frac{\delta {\cal G}}{\delta \Psi^*} + \zeta_{\Psi}(\vec r,t),
\label{eq1} \\ \Gamma^{-1}_{A}\frac{\partial \vec A}{\partial t}&=& -
\frac{\delta {\cal G}}{\delta \vec A} + \vec \zeta_{A}(\vec r,t)
\label{eq2}, \end{eqnarray} 
where $\Psi$ is the order parameter and
$\vec A$ the vector potential. $\Gamma_{\Psi}=2mD/\hbar^2$ and
$\Gamma_{A}=c^2/\sigma_{n}$ are kinetic coefficients where $\sigma_{n}$
is the quasiparticle conductivity and $D$ is the electron diffusion
constant. $\zeta_{\Psi}(\vec r,t)$ and $\vec \zeta_{A}(\vec r,t)$ are
Langevin thermal noise with average $\langle \zeta_{\Psi}
\rangle = \langle \zeta_{A}  \rangle=0$ and the
following correlations \cite{enomoto,hohenberg}: 
\begin{eqnarray}
\langle \zeta_{\Psi}^*(\vec r,t)\zeta_{\Psi}(\vec r',t')\rangle= 2 k_B
T\Gamma^{-1}_{\Psi}\delta (\vec r - \vec r')\delta(t -t'), \label{eq3}
\\ \langle \zeta_{A}^i(\vec r,t)\zeta_{A}^j(\vec r',t') \rangle= 2 k_B
T\Gamma^{-1}_{A}\delta (\vec r - \vec r')\delta(t -t')\delta_{ij}.
\label{eq4} \end{eqnarray}

The difference $\cal{G}$ between the superconducting and the
normal state Gibbs free energy is given by
\begin{eqnarray}
{\cal G}&=&{\cal G}_s-{\cal G}_n \nonumber \\
&=&\int \Bigl[ \alpha(T)|\Psi|^2+\frac{1}{2}\beta|\Psi|^4
+\frac{1}{2m_s}|(\frac{\hbar}{\imath}\nabla-\frac{e_s}{c}\vec{A})
\Psi |^2  \nonumber \\
&+&\frac{1}{8 \pi}(\nabla \times A-H_a)^2 \Bigr] d^3\vec r \label{eq5}.
\end{eqnarray}
To include demagnetization fields, the magnetic term in $\cal{G}$
(last term) must be integrated inside and outside the sample
volume. The differential Eqs. (\ref{eq1}) and (\ref{eq2}) must be
solved in the whole space.

Taking the variations in Eqs. (1) and (2), the normalized TDGL equations become
\cite{enomoto,gropp,kato,HD1,HD2}
\begin{eqnarray}
\frac{\partial \Psi}{\partial t}&=&\frac{1}{\eta}
[(1-T)\Psi(1-|\Psi |^2)  \nonumber \\
&-&(-i \nabla -{\vec A})^2 \Psi ]+ \frac{\zeta_{\Psi}(\vec r, t)}{\sqrt{1-T}}, \label{eq6} \\
\frac{\partial {\vec A}}{\partial t}&=&(1-T)\mbox{Im}[\Psi^* (
\nabla - i{\vec A})\Psi] \nonumber \\
&-&\kappa^2 \nabla \times \nabla \times \vec A + \vec \zeta_{A}(\vec r,t). \label{eq7}
\end{eqnarray}
Lengths have been scaled in units of the coherence length
$\xi(0)$, times in units of $t_0=4\pi \sigma_n \lambda_L^2/c^2$,
$\vec{A}$ in units of $H_{c2}(0) \xi (0)$, $\Psi$ in units of
$\Psi_{\infty}=[mc^2/8\pi e^2\lambda(T)^2]^{1/2}$ and temperature in units
of $T_c$. $\eta$ is equal to the ratio of the characteristic time
$t_0$ for the relaxation of $\vec{A}$ and the time $t_{GL}$ for
the relaxation of $\Psi$: $\eta=t_{GL}/t_0=c^2/(4\pi \sigma_n
\kappa^2 D)$, with $t_{GL}=\xi^2/D$.
For superconductors with magnetic impurities we have
$D_{imp}=c^2/(48\pi \sigma_n \kappa^2)$, and therefore $\eta=12$
in this case.

The normalized thermal noise correlations are
\begin{eqnarray}
\langle \zeta_{\Psi}^*(\vec r,t)\zeta_{\Psi}(\vec r',t') \rangle=
\frac{8\pi}{\eta} \sqrt{2G_i} T \delta (\vec r - \vec r')\delta(t -t'), \label{eq8}\\
\langle \zeta_{A}^i(\vec r,t)\zeta_{A}^j(\vec r',t') \rangle= 4
\pi \sqrt{2G_i} T\delta (\vec r - \vec r')\delta(t -t')\delta_{ij}
\label{eq9},
\end{eqnarray}
where
\begin{equation}
G_i=\frac{1}{2}\left[\frac{k_BT_c}{H_c^2(0)\xi^3(0)}\right]^2 \label{Gi}
\end{equation}
is the Ginzburg number \cite{ginzburg,blatter} that governs the
strength of thermal fluctuations. $G_i$ measures the relative size of
the minimal (T=0) condensation energy $H_c^2(0)\xi^3(0)$ within a
coherence volume and the energy of thermal fluctuations at $T_c$,
$k_BT_c$. For conventional low-$T_c$ superconductors $G_i \approx
10^{-8}$ and for high-$T_c$ superconductors $G_i \approx 10^{-4}$ (see
Ref. \onlinecite{blatter}).

Equation (\ref{eq7}) is Ampere's law, i.e.
\begin{eqnarray}
\vec J=\kappa^2 \nabla \times \nabla \times \vec A \label{eq10},
\end{eqnarray}
where $\vec J$ is the current inside the sample:
\begin{eqnarray}
\vec J(\vec r)&=&\vec J_n(\vec r) + \vec J_s(\vec r) + \vec \zeta_{A}(\vec r, t), \label{eq11} \\
\vec J(\vec r)&=&-\frac{\partial {\vec A}}{\partial t}+(1-T)\mbox{Im}[\Psi^* ( \nabla - i{\vec
A})\Psi] \nonumber
\\
&+& \vec \zeta_{A}(\vec r,t). \label{eq12}
\end{eqnarray}
$J_n$ is the current due to the normal electrons, $\vec
J_n=\sigma_{n} \vec E$, and $J_s$ is the supercurrent.

The Biot-Savart law relates the magnetic induction ${\vec B}=\nabla \times {\vec A}$, the applied
magnetic field $\vec H_a$ and the sample currents:
\begin{eqnarray}
{\vec B}({\vec r})-{\vec H}_{a}&=& \frac{1}{4\pi \kappa^2} \int {\vec J} ({\vec {r'}}) \times
\frac{{\vec r}-{\vec {r'}}}{|{\vec r}-{\vec {r'}}|^3}d^3{\vec {r'}}, \label{eq13} \\
\vec B({\vec r})-\vec H_a&=&\int_A Q({\vec r},{\vec {r'}}) g({\vec {r'}}) d^3{\vec {r'}}
\label{eq14}.
\end{eqnarray}
This equation states that the magnetic field is completely determined
by the externally applied field $H_a$ and by the currents flowing
inside the sample (which are given by $ g(\vec {r'})$). The scalar
function $g(\vec {r'})$ is the local magnetization or density of tiny
current loops and is given by \cite{brandt}
\begin{eqnarray}
\vec{J}(x,y)=-\hat{\bf z}\times \nabla g(x,y)=\nabla\times\hat{\bf z} g(x,y). \label{eq15}
\end{eqnarray}
This guarantees that ${\nabla \cdot \vec J}=0$.

The kernel $Q$ can be obtained from the Biot-Savart law. In the
dipolar approximation, i.e. for long distances, ${\bf \rho}=|{\vec
r}-{\vec {r'}}|$ and $Q=-1/4\pi\rho^3$.

Then $g(\vec r)$ and $Q(\vec r, \vec {r'})$ have all the information
that we need to include the demagnetization fields (and $Q(\vec r, \vec
{r'})$ is a known and time-independent kernel). If we write the TDGL
equations in terms  of $g(\vec r)$, and we use the Biot-Savart law,
we only need to solve the differential equations inside the sample
volume and the demagnetization fields are automatically included.\cite{HD04} 
On the other hand, we can also solve Ampere's law
retaining the 3D nature of the magnetic field distribution. Both
methods are equivalent for steady-state magnetic phenomena ($\nabla
\cdot J=0$) when the Biot-Savart and Ampere's laws are equivalent.
\cite{jackson}

In the present paper we will deal with square samples with thickness
$d$ ($d<\lambda$). In this case, the order parameter and $g(x,y)$ can
be assumed to be uniform in the $\hat{\bf z}$ direction. Using the
Biot-Savart law it is only necessary to write the TDGL equations for a
2D sample,\cite{HD04} i.e.
\begin{eqnarray}
\frac{\partial \Psi}{\partial t}&=&\frac{1}{\eta} [(1-T)\Psi(1-|\Psi |^2) - (-i \nabla_{2D}
-{\vec A})^2 \Psi ]
\nonumber \\
&+& \frac{\zeta_{\Psi}(\vec R, t)}{\sqrt{1-T}}, \label{eq18} \\
\frac{\partial {\vec A}}{\partial t}&=&(1-T)\mbox{Im}[\Psi^* (
\nabla_{2D} - i{\vec A})\Psi]-\nabla \times \hat{\bf z}g \nonumber \\
&+& \vec\zeta_{A}(\vec R,t), \label{eq19} \\
B_z(x,y)&-&H_a=\int_A Q({\vec R},{\vec {R'}}) g({\vec {R'}}) d^2{R'}, \label{eq20}
\end{eqnarray}
were $\vec R=(x,y)$ and $\vec {R'}=(x',y')$. We assumed $H_a || \hat{\bf z}$. To obtain $g(x,y)$ we
invert Equation (\ref{eq20}) at each time step using the efficient Conjugate Gradient
Method.\cite{daniel}

The boundary conditions for this problem, at the surface of the
superconducting sample, are
\begin{eqnarray}
(-i\nabla_{2D} -\vec{A})\Psi|_n =0, \label{eq21} \\
{\mathbf g}|_b=0. \label{eq22}
\end{eqnarray}

\subsection{Stationary Ginzburg-Landau theory}

To calculate the equilibrium states we minimize the mean-field
free energy without thermal fluctuations and retain the
three-dimensional magnetic field distribution. The system of GL
equations, using dimensionless variables and the London gauge
$\nabla \cdot \vec{A}=0$, has the following form:
\cite{PRB57v,PRL81v,PRB65}
\begin{eqnarray}
\left( -i\nabla_{2D}-\vec{A}\right) ^{2}\Psi
=\Psi \left( 1-\left| \Psi \right| ^{2}\right), \label{eq23} \\
-\Delta _{3D}\vec{A}=\frac{d}{\kappa ^{2}}\delta \left( z\right) \vec{j}_{2D} \label{eq24},
\end{eqnarray}
where
\begin{eqnarray}
\vec{j}_{2D}=\frac{1}{2i}\left( \Psi ^{\ast }\nabla_{2D}\Psi -\Psi \nabla_{2D}\Psi
^{\ast }\right) -\left| \Psi \right| ^{2}\vec{A}, \label{eq25}
\end{eqnarray}
is the density of superconducting current. The order parameter satisfies the boundary conditions:
\begin{eqnarray}
\left. \left( -i\nabla_{2D}-\vec{A}\right) \Psi \right| _{n}=0. \label{eq26}
\end{eqnarray}
The boundary condition for the vector potential in cartesian coordinates becomes:
\begin{eqnarray}
\vec{A} |_{|x|=R_{s},|y|=R_{s}}=H_{0}(x,-y)/2, \label{eq27}
\end{eqnarray}
at the boundary $R_s$ of a larger space grid.

Here the distance is measured in units of the coherence length $\xi $,
the vector potential in $c\hbar/2e\xi$, and the magnetic field in
$H_{c2}=c\hbar/2e\xi ^{2}=\kappa \sqrt{2}H_{c}$. The superconductor is
placed in the $(x,y)$ plane, the external magnetic field is directed
along the $z$ axis, and the indices 2D, 3D refer to two- and
three-dimensional operators, respectively.

\section{Results}

We will compare the results obtained from minimization of the GL free
energy (with magnetic fields solved via Ampere's law) with the
stationary states obtained from the dynamics of the TDGL equation (with
magnetic fields solved via Biot-Savart law). We study the influence
of the square symmetry restrictions on the vortex entrance in square
samples. In a first step we will neglect the thermal fluctuations and
we will investigate vortex states in mesoscopic superconducting squares
at $T = 0$ and $T = 0.7T_c$, both within the time-dependent GL theory
and within the stationary GL theory. In a second step we will take into
account the thermal fluctuations in the TDGL theory and again we will
compare the obtained results with the ones from the stationary GL
theory. Such a comparison will be performed both for conventional
low-$T_c$ and for high-$T_c$ superconductors. In the last part we focus
on the time average of the order parameter $|\Psi|^2_m$ and the
number of vortices $N_v$ for low and high-$T_{c}$ superconductors.

\subsection{No thermal fluctuations}

\begin{figure}[htb]
\includegraphics[width=1.0\linewidth]{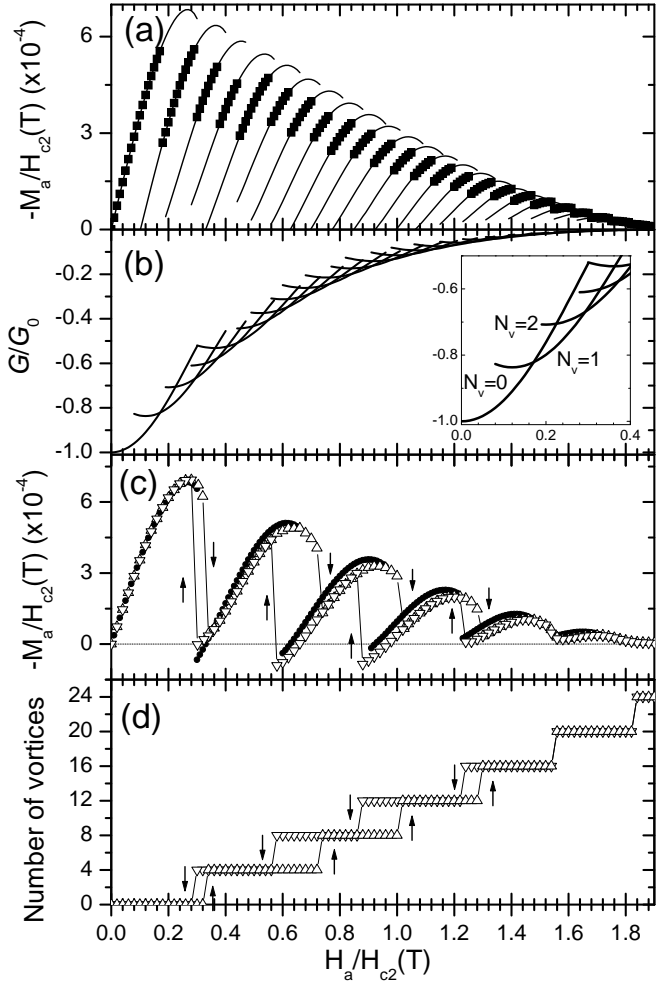}
\caption{\label{fig1} Results at $G_i=0$ and $T=0$. (a) The
squares (${\scriptstyle \blacksquare}$) show the apparent
magnetization corresponding to the energy ground state and the
straight lines the $M_a$ results of all the meta-stable states. (b)
Gibbs free energy of the different vortex states, the inset is a
zoom at low fields. Apparent magnetization (c) and the number of
vortices (d) calculated with the TDGL equations, when increasing
($\vartriangle$) and decreasing ($\triangledown$) the external
magnetic field. The arrows show the vortex transition with TDGL.}
\end{figure}

First we consider superconducting squares with $G_i=0$, for which,
thermal fluctuations are not taken into account. Fig.~\ref{fig1}
shows our results as a function of the applied magnetic field for
a superconducting square with sides equal to $W = 10\xi(0)$ and
thickness $d = 0.1\xi(0)$. The GL parameter equals $\kappa = 1$
and temperature $T = 0$.

Throughout the paper, we will show results of the apparent
magnetization defined as $4 \pi M_a=\langle B_z \rangle -H_a$. The
real magnetization is equal to $4 \pi M=\langle B_z \rangle -H$
where H is the internal magnetic field that can also be written as
$H=H_a-(4\pi M)N$ where N is the demagnetization factor that
depends on the sample geometry. However, $M_a$ is what is measured
 experimentally.

The square symbols in Fig.~\ref{fig1}(a) give the results of the
apparent magnetization, corresponding to the ``ground state" (global
minimum of the mean-field free energy) as calculated within the stationary GL
theory. The ground state vortex transitions are found calculating
the free energy of all the possible (meta-stable) vortex states
that are shown in Fig.~\ref{fig1}(b). The inset is a zoom of the
free energy at low fields. The different branches correspond to
different values of the total vorticity, 
$N_v= \frac{1}{\Phi_o}\oint (A+\frac{J_s}{|\Psi|^2})dl$. 
Then, according to the
global minima of the free energy, the vorticity changes always by
one unit ($\Delta N_v=1$).

\begin{figure}
\includegraphics[width=1.0\linewidth]{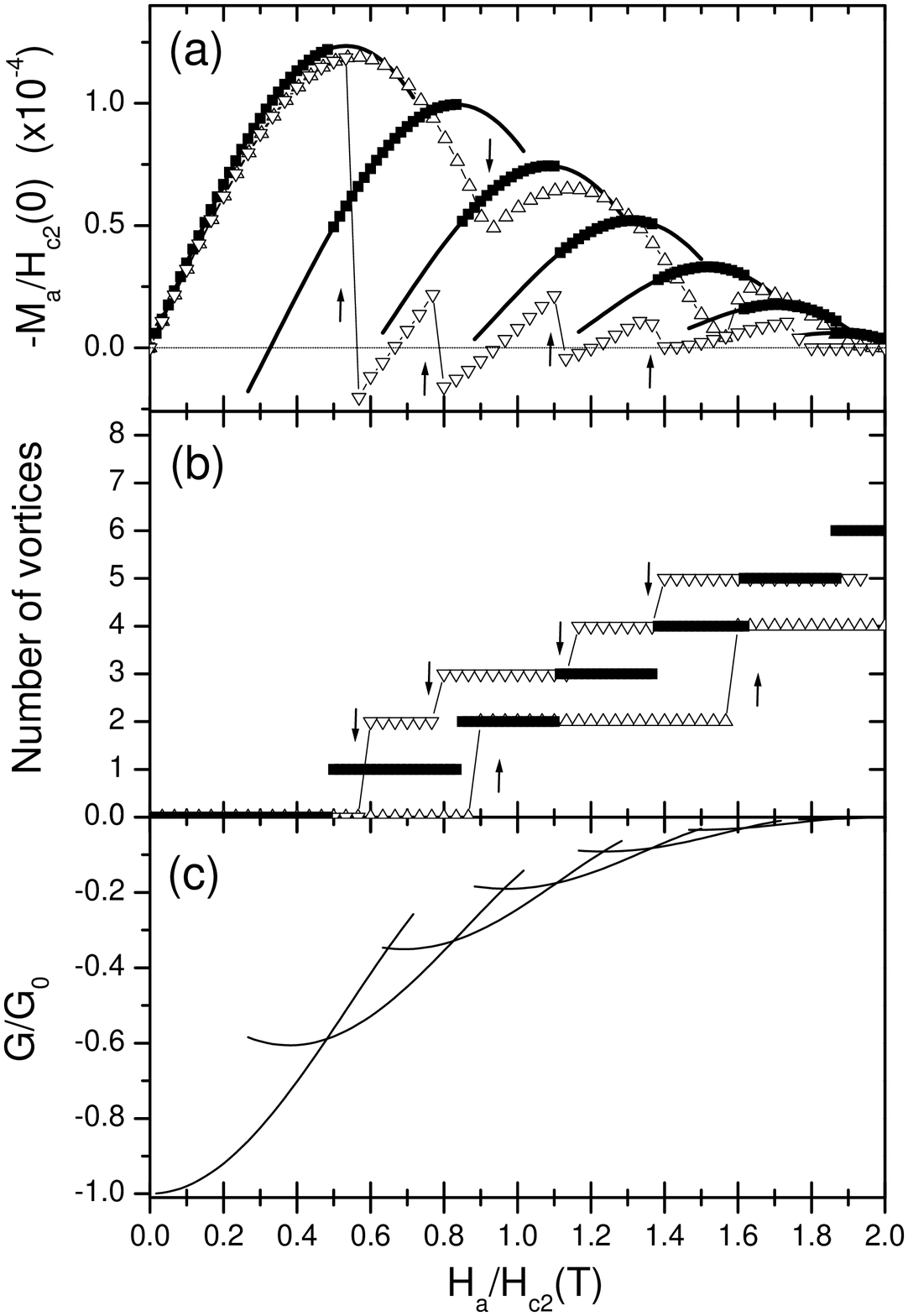}
\caption{\label{fig2} Similar results as Fig.~\ref{fig1} but now
at $T=0.7T_c$ and $G_i=0$. 
(a) Apparent magnetization, $M_a$.
The results obtained  minimizing the mean-field
GL free energy (${\scriptstyle
\blacksquare}$) are compared to the 
dynamical results calculated with the TDGL equations,
when increasing
($\vartriangle$) and decreasing ($\triangledown$) the external
magnetic field. 
(b) The total vorticity vs. $H_a$ for the same cases as in (a).
(c) Gibbs free energy of the different vortex states as a function of $H_a$.}
\end{figure}

In Fig.~\ref{fig1}(c) we study the behavior of the apparent
magnetization within the framework of the TDGL theory. The
triangles show the behavior for increasing ($\vartriangle$) and
decreasing ($\triangledown$) field. Notice that the transitions
while increasing the field or decreasing the field do not occur at
the same field values, i.e. the penetration field ($H_p$) differs
from the expulsion field ($H_e$). This hysteresis is due to the
surface barrier.\cite{PRB59, HD1} With increasing and decreasing
field we see that the vorticity changes by four and minus four at
the transitions, which means that the vortices enter or leave the
sample by four, i.e. one vortex through each side of the
square. This is shown more explicitly in Fig.~\ref{fig1}(d) where
the total vorticity, {\it i.e.} the number of vortices inside the
sample, are plotted as a function of the applied magnetic field. Of course,
following the ground state free energy, the vorticity changes
always by one unit. 
The fact that in the TDGL results the vorticity changes by four at 
the transitions means that, in this case, there is not a simple dynamical 
path to break the square symmetry in the system during the field sweep.
The close circles ($\bullet$) in Fig.~\ref{fig1}(c) are the
$-M_a/H_{c2}(T)$ values corresponding to the meta-stable states
with vorticity equal to a multiple of four as calculated within
the stationary GL theory. Notice that the values of
$-M_a/H_{c2}(T)$ obtained using the TDGL and stationary GL theory
are similar for the same number of vortices.

Next, we investigate the influence of finite temperature
neglecting the thermal fluctuations, i.e. $G_i=0$.
Figs.~\ref{fig2}(a-c) show the same as Figs.~\ref{fig1}(b-d), but
now at temperature $T=0.7 T_c$. When we increase the temperature,
both $\xi(T)=\xi(0)/\sqrt{1-T/T_c}$ and
$\lambda(T)=\lambda(0)/\sqrt{1-T/T_c}$ increase and the size of
the sample decreases measured in units of $\xi(T)$,
($\xi(T=0.7T_c)/\xi(0)=1.83$). In this case less vortex
penetration events are necessary to arrive to the normal state.
Notice that this is what happens in a typical experimental
situation when the temperature is increased. 
The main difference with the zero-temperature situation is that 
in the present case the vorticity changes with $\Delta N_v= 2$ for increasing 
magnetic field while the vorticity changes with $\Delta N_v= - 1$ in most transitions 
for decreasing $H_a$. This behavior is different from the results obtained in Fig. 1(d) because 
of the small size of the sample as compared to $\xi(T)$. 
For example, in Fig.~\ref{fig2}(b) we plot the number of vortices for increasing 
($\vartriangle$) and decreasing ($\triangledown$) the magnetic field and we
compare these results with the vorticity of the energy ground
state (${\scriptstyle \blacksquare}$). The arrows show the vortex
transitions with TDGL. In Fig.~\ref{fig2}(b) we see the presence
of hysteresis due to meta-stable states.

\subsection{Thermal fluctuations}

Next, we include thermal fluctuations, i.e. $G_i \neq 0$ in Eq. (\ref{Gi}).

\begin{figure}
\includegraphics[width=1.0\linewidth]{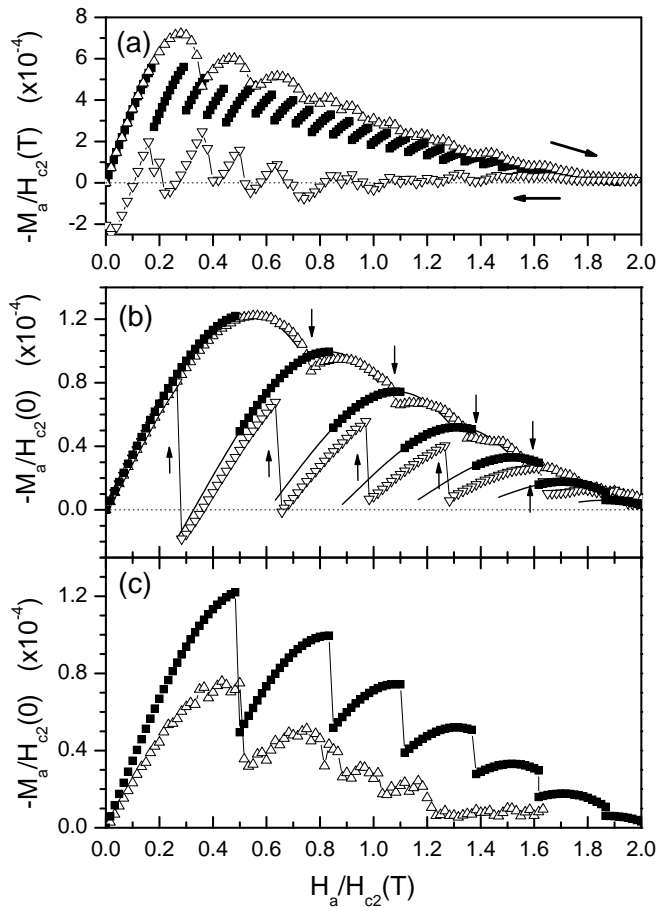}
\caption{\label{fig3} Results obtained including thermal
fluctuations for $T=0.7T_c$. (a) $G_i=10^{-10}$, (b) $G_i=10^{-8}$
and (c) $G_i=10^{-5}$. The system size in (a) is $W=10 \xi(T)$,
$d=0.1\xi(T)$ and in (b) and (c) $W=10 \xi(0)$, $d=0.1\xi(0)$. The
triangles are for increasing ($\vartriangle$) and decreasing
($\triangledown$) the magnetic field and the solid squares are the
results corresponding to the mean field ground states.}
\end{figure}

It is known that the free energy minima results obtained at
temperature $T_1$ can be rescaled and will be equal to the free
energy minima results of an ``equivalent" system at a different
temperature $T_2 \neq T_1$. To be equivalent the systems must have
the same size measured in units of $\xi(T)$ and the same $\kappa$.
In Fig.~\ref{fig3}(a) we compare the ground state apparent
magnetization results for two situations: (i) a square with sides
$W = 10\xi(0)$ and thickness $d = 0.1\xi(0)$ at $T = 0$, and (ii)
for a square with $W = 10\xi(T)$ and $d = 0.1\xi(T)$ at $T
=0.7T_c$ for $G_i =10^{-10}$. As the size of the system is the
same measured in $\xi(T)$ the latter results can be compared if we
normalize the field by the temperature dependent $H_{c2}(T)$, as
we did in Fig~\ref{fig3}(a). We see that thermal fluctuations
have broken the symmetry restrictions for vortex entrance and in this
case $\Delta N_v=2$ for $H_a<0.8H_{c2}(T)$ and $\Delta N_v=1$ for
$H_a>0.8 H_{c2}(T)$, while $\Delta N_v=4$ when neglecting the
thermal fluctuations (see Fig.~\ref{fig1}(a)). Even when $\Delta
N_v=1$, the dynamic TDGL results for $-M_a$ (open symbols) do not
follow in detail the $-M_a$ curves obtained from free energy
minimization (solid symbols). This behavior will be more clear in
Fig.~\ref{fig3}(b) in case of a smaller sample.

In Fig.~\ref{fig3}(b) we show the apparent magnetization as a
function of the applied magnetic field for a square with sides
equal to $W = 10\xi(0)$ and thickness $d = 0.1\xi(0)$ at $T
=0.7T_c$ for $G_i =10^{-8}$, i.e. a conventional low-$T_c$
superconductor. The square symbols and the solid curves are the
apparent magnetization for the ground state and the meta-stable
states as calculated within the framework of the stationary
Ginzburg-Landau theory. The triangles indicate the results with
increasing ($\vartriangle$) and decreasing ($\triangledown$) field
when using the time-dependent Ginzburg-Landau theory, taking into
account the thermal fluctuations. For this system size and
$G_i=10^{-8}$ the TDGL results with increasing and decreasing
magnetic field show that $\Delta N_v=1$. But, even though $\Delta
N_v=1$, there is still hysteresis in the $-(M_a)$ curve for such a
conventional mesoscopic low-$T_c$ superconductor. The system
clearly does not follow the mean field free energy minima behavior. Notice
that there is a rather good agreement between the results from the
TDGL theory and the stationary GL theory, both for the values of
$-M_a$ and for the transition fields.

Even when the thermal fluctuations are weak, as in
Fig.~\ref{fig3}(b), it is expected that
 for a sufficient long simulation time the apparent magnetization would relax until the ground
state is reached. In this case, the relaxation of the magnetic
flux is related with the possibility that vortices overcome the
surface barrier by thermal activation. In our simulations we allow
the system to relax during 250000 time steps with $\Delta t=0.0045
t_o$, which gives the system the possibility to relax during
$t=1125 \;t_{0} \sim 0.1 \mu \mbox{sec}$. However, the hysteresis
in the apparent magnetization curve is also observed
experimentally \cite{GN390} in mesoscopic low-$T_c$
superconductors. Consequently, we can conclude that the half-life
time of the meta-stable states is very large in low-$T_c$
superconductors even in comparison with the usual experimental
times.

Next, we repeat the calculation for the same sample as in
Fig.~\ref{fig3}(b), but now including thermal fluctuations of strength
corresponding to a high-$T_c$ superconducting material ($G_i
=10^{-5}$). The results are shown in Fig.~\ref{fig3}(c). We see that
the first two penetration fields obtained from the TDGL equations
agree very well with  the ones resulting from the mean-field free energy minima. 
This means that the effect of the
surface barrier is almost suppressed  here, 
once we allow the system to relax a short time.
On the other hand, 
the strong thermal fluctuations decrease
the $-(M_a)$ values in Fig.~\ref{fig3}(c), with
respect to the mean-field values. 
Below the first penetration field,
in the Meissner state, the magnetic flux inside the sample
($BW^2$) depends only on the penetration length. 
Therefore, in the Meissner
branch of figure Fig.~\ref{fig3}(c) we can deduce  from the
slope of $M_a(H_a)$  what
appears to be an increase in $\lambda$ induced by thermal
fluctuations. In fact, it is known that
thermal fluctuations can induce 
an increase of the effective London penetration length
in high-$T_c$ superconductors (see for example Ref.~\onlinecite
{lobb}), which is a consequence
of the renormalization of the superfluid  density due
to thermal fluctuations, since
$\lambda_{\mbox{eff}}^2 = mc^2/(8\pi e^2 \langle |\Psi|^2\rangle)$.

\begin{figure}
\includegraphics[width=1.0\linewidth]{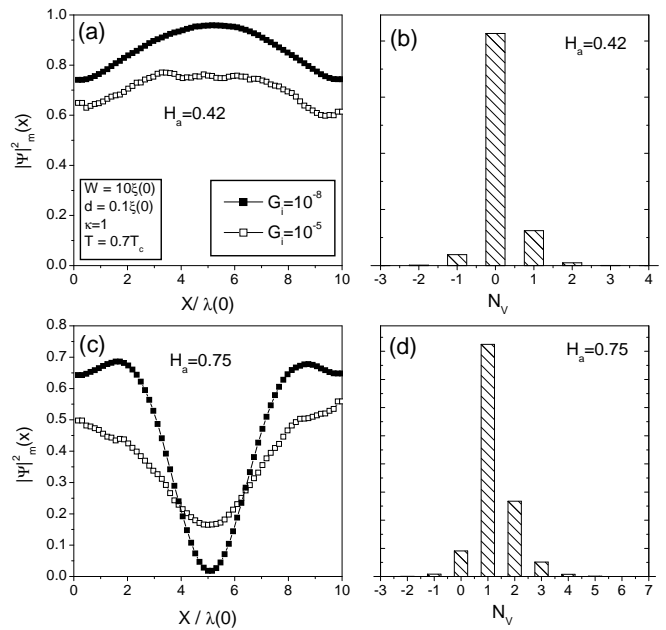}
\caption{\label{fig4} Time average of the order parameter in the
$x$ direction, $|\Psi|^2_m$, for $G_i=10^{-8}$ (${\scriptstyle
\blacksquare}$) and $G_i=10^{-5}$ (${\scriptstyle \square}$) for
(a) $H_a=0.42 H_{c2}(T)$ and (c) $H_a=0.75 H_{c2}(T)$. (b) and (d)
are histograms of the different vortex states obtained when
$G_i=10^{-8}$.}
\end{figure}

In order to understand these effects in Fig.~\ref{fig3}(c), we investigate the
time average of the order parameter $\langle|\Psi|^2\rangle$ and the time
variations in the number of vortices ($N_v(t)$) for low and high-$T_{c}$ 
superconductors, with $G_i=10^{-8}$ and $G_i=10^{-5}$,
respectively. In Figs.~\ref{fig4}(a) and \ref{fig4}(c) we show
$\langle|\Psi|^2\rangle$ along the x direction, i.e. from the middle of one
side to the middle of the opposite side of the square, for two
values of the applied magnetic field, $H_{a} = 0.42H_{c2}(T)$ and
$H_{a} = 0.75H_{c2}(T)$. We obtain the following results: for
$G_i=10^{-8}$ the number of vortices do not change in time and the
$\langle|\Psi|^2\rangle(x)$ results (${\scriptstyle \blacksquare}$) are near
to the mean field ones ($G_i=0$). However, for $G_i=10^{-5}$ we
observe that $\langle|\Psi|^2\rangle(x)$ (${\scriptstyle \square}$) is
lower than the mean field (MF) result and the number of
vortices fluctuates in time.

The field $H_a=0.42H_{c2}(T)$, 
corresponds to the Meissner 
state without vortices in the mean-field
case,  see Fig.~\ref{fig3}(b). For $G_i=10^{-8}$
we also obtain that $N_v(t)\equiv 0$ for all $t$. However, for
$G_i=10^{-5}$ the number of vortices change in time. In
Fig.~\ref{fig4}(b) we plot a histogram of the different vortex
states obtained during 250000 time steps. We see that the state
with $N_v=0$ is the most probable one, but that there is also some
 probability to have jumps to other vortex states, mainly with $N_{v}=\pm 1$.
Thus the reduction of $\langle|\Psi|^2\rangle$ from the mean field result 
for $G_i=10^{-5}$ is a consequence of thermal fluctuations which allow 
the nucleation of thermally induced vortex-antivortex pairs 
\cite{KT,Beasley,Hebard,Kadin} as well as the entrance and exit of thermally activated 
vortices. This leads to an increase of the effective $\lambda$, due to the lowering of 
$\langle|\Psi|^2\rangle$.

Results similar to Figs.~\ref{fig4}(a)-(b) are obtained in
Figs.~\ref{fig4}(c)-(d) for $H_{a} = 0.75H_{c2}(T)$.
Fig.~\ref{fig4}(c) shows that, for $G_i=10^{-8}$ (${\scriptstyle \blacksquare}$), we have a vortex
state with one vortex: $\langle|\Psi|^2\rangle=0$ in the center of the sample
where the vortex is located. However, for $G_i=10^{-5}$(${\scriptstyle \square}$), we
observe that $\langle|\Psi|^2\rangle(x)$ is different from zero in the center
of the sample. The corresponding vortex state histogram is shown
in Fig.~\ref{fig4}(d). We see that the state with one vortex is
the most probable one but that there is also a finite probability
for other vortex states which contribute differently to
$\langle|\Psi|^2\rangle$. This makes
$\langle|\Psi|^2\rangle \neq 0$  in the center of the sample. 
There are continuous entrances and exits of
vortices and the time average is a mixture of states with different vorticity.

More details about thermal activation of vortices can be obtained
by calculating the average number of vortices $\langle
N_v\rangle$:
\begin{equation}
\langle N_v\rangle= \frac{\Delta t}{t_f-t_i} \sum_{t=t_i}^{t_f} N_v(t)\;\label{v-medio},
\end{equation}
and the probability to have $N_v=n$ vortices ($P_{(N_v=n)}$):
\begin{equation}
P_{(N_v=n)}= \frac{\Delta t}{t_f-t_i} \sum_{t=t_i}^{t_f}
\delta_{N_v(t);n}\;\label{v-probab},
\end{equation}
both magnitudes can be calculated if we know the number of
vortices in each time step $N_v(t)$. Where the sum in Eqs.
(\ref{v-medio}) and (\ref{v-probab}) are taken between two time
values ($t_i$ and $t_f$), and $\delta_{N_v(t);n}$ is the Kronecker
delta.

\begin{figure}
\includegraphics[width=1.0\linewidth]{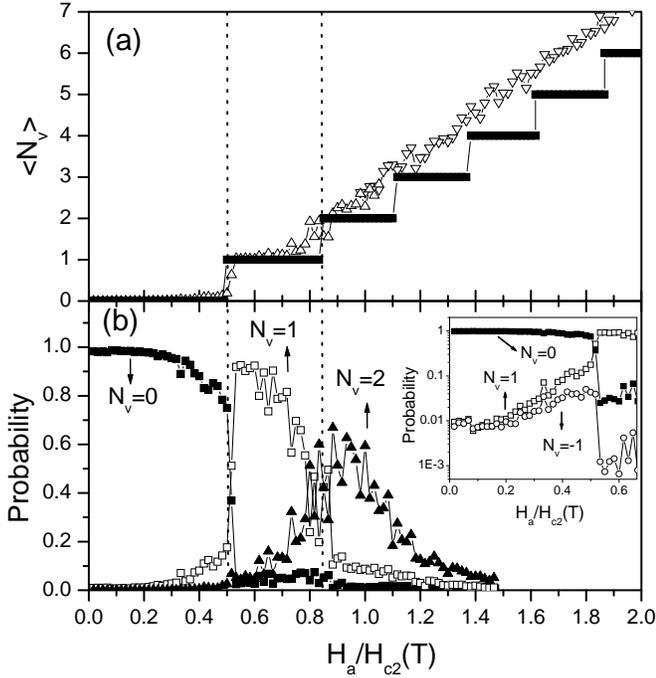}
\caption{\label{fig5} (a) Time average of the different vortex
states obtained for $G_i=10^{-5}$ for increasing ($\vartriangle$)
and decreasing ($\triangledown$) the magnetic field. The black
squares (${\scriptstyle \blacksquare}$) are the ground state
results. (b) Probability of the different vortex states. The inset
is a zoom at low fields in a logarithmic scale. }
\end{figure}

In Fig.~\ref{fig5}(a) we plot the average number of vortices and
in Fig.~\ref{fig5}(b) the probabilities to find a specific vortex
state as a function of the applied magnetic field. These results
where obtained during the magnetic field sweep up shown in
Fig.~\ref{fig3}(c) ($G_i=10^{-5}$).

In Fig.~\ref{fig5}(a), for $G_i=10^{-5}$ ($\vartriangle$) and
small magnetic fields ($H_a<0.7H_{c2}(T)$), we see that $\langle
N_v \rangle$ vs. $H_a$
 follows the mean field result ($N_v ^{MF}(H_a)$) (${\scriptstyle \blacksquare}$).
For $H_a>0.7H_{c2}(T)$ the results are different, except at the start of
each step where $\langle N_v \rangle=N_v ^{MF}$. For strong
thermal fluctuations we see that the average number of vortices
can change continuously and is frequently larger than $N_v ^{MF}$.

More details about the thermal activation of vortices can be
obtained calculating the probability to have $N_v$ vortices using
Eq. (\ref{v-probab}). In Fig.~\ref{fig5}(b) we show the
probabilities of the different vortex states for $G_i=10^{-5}$.
The inset is a zoom of the low magnetic field region in a
logarithmic scale. Near $H_a=0$, the probability of the zero
vortex state is near one ($P_{(N_v=0)} \approx 1$) and we also see
that $P_{(N_v=1)}\approx P_{(N_v=-1)}\approx 0.01$. From the inset
we find that for $H_a \approx 0.5H_{c2}(T)$ there is a discontinuous jump
down in $P_{(N_v=0)}$ and that at the same time the probability of
having $N_v=1$ increases abruptly. When this happens we see that
$\langle N_v \rangle=1$  in Fig.~\ref{fig5}(a).

We also calculated the fluctuations in the vorticity ($N_v(t)$)
and the fluctuations in $-M_a(t)/H_{c2}(0)$. The fluctuations in
the magnitude of the apparent magnetization $M_a(t)$ is obtained
through the relation:
\begin{equation}
\Delta M_a= \sqrt{\langle M_a^2\rangle-\langle M_a\rangle^2} \;\;,
\end{equation}
where the mean values of $M_a$ (i.e. $\langle M_a\rangle$ and
$\langle M_a^2\rangle$) are calculated using the definition given
in Eq. (\ref{v-medio}).

\begin{figure}
\includegraphics[width=1.0\linewidth]{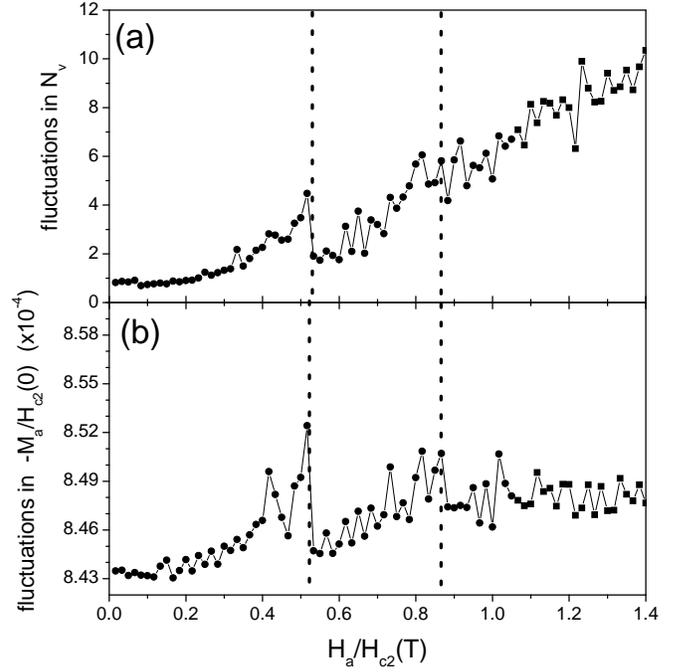}
\caption{\label{fig6} (a) Fluctuations in the number of vortices
and (b) in $-M_a/H_{c2}(0)$ for $G_i=10^{-5}$ corresponding to the
situation of Fig.~\ref{fig5}. }
\end{figure}

The fluctuations are shown in Fig.~\ref{fig6}. From
Fig.~\ref{fig6}(a) we see that the fluctuations in the vorticity
increase with increasing field until the first penetration
field, $H_{p1}$, (first dashed vertical line), where
there is a remarkable decrease in the number of vortex fluctuations. 
The fluctuations in $M_a$
also decrease at $H_{p1}$ as we can see from Fig.~\ref{fig6}(b).
At the second penetration
field (second dashed line) in Fig.~\ref{fig6} the decrease in the
fluctuations at $H_{p2}$ is almost completely buried under the
thermal noise due to a continuous increase in the fluctuations
of the number of vortices with increasing field, as seen in
Fig.~\ref{fig6}(a).

\section{Conclusions}

We studied the influence of the strength of thermal fluctuations
on the magnetic behavior of mesoscopic superconductors. In the
absence of thermal fluctuations and surface imperfections, we
found that the vorticity changes by four at the transition fields
due to the symmetry of the mesoscopic square sample. The surface
barriers and the symmetry of the sample produce hysteresis and
meta-stable states.

In low-$T_c$ mesoscopic superconductors, where thermal
fluctuations are small, we did not find thermally activated
entrance/exit of vortices through surface barriers. This result
agrees with experiments in low-$T_c$ superconductors where
hysteresis in the penetration fields and meta-stable states were
found.

In low-$T_c$ superconductors, small thermal fluctuations however
are enough to
break the square symmetry restriction to vortex entrance and less
vortices penetrate at the same field.

A different behavior was found when the strength of thermal
fluctuations is increased as is the case for high-$T_c$
superconductors. In this case there are frequent thermally
activated events of
entrance/exit of vortices in agreement with recent experimental
results on micron-sized high-$T_c$ superconducting thin film rings
\cite{kogan-clem}. We also found that thermal fluctuations
increase the effective London penetration depth if we compare it with the
mean field result.

In high temperature superconductors,  
the $d$-wave symmetry of the ground state and the
tetragonal symmetry of the underlying crystalline 
structure can have an important effect in some of their
macroscopic properties. These have 
been modeled by phenomenological Ginzburg-Landau or London theories, 
containing  mixed gradient couplings to an order parameter with a 
different symmetry \cite{berlinsky,heeb}
or additional quartic derivative nonlocal terms.\cite{ichioka,kogan}
An important consequence of this is that, at large 
magnetic fields, the equilibrium vortex lattice deviates from the 
triangular Abrikosov lattice to a square lattice, as it has been
observed recently.\cite{brown} Also the  possible effects in
the individual  structure of vortices in cuprates superconductors 
has been the subject of intense research.\cite{hoffman}
In the case of mesoscopic high-$T_c$ superconductors, 
the structure of the  vortex
configurations confined within the geometry of the sample 
and  the  quantitative values of the  surface barrier
and penetration fields could be affected by
such considerations. On the other hand, 
the  qualitative physics of
the {\it thermally activated} entrance/exit of 
vortices through surface barriers should remain similar to
the one described here with the dynamics of the simple GL equations. 
In any case, more 
experimental and theoretical works in this field are necessary.
Therefore, the study of the physical properties of mesoscopic high-$T_c$ 
superconductors can contribute to the global understanding of these 
interesting materials.

\section{Acknowledgments}

This work was supported by the Argentina-Belgium collaboration programme
SECYT-FWO FW/PA/02-EIII/002.
DD and ADH acknowledge support from ANPCYT PICT99-03-06343, from
CONICET and from CNEA P5-PID-93-7.
FMP and BJB acknowledge support from
 the Flemish Science Foundation (FWO-Vl) and the
Belgian Science policy. B.~J.~Baelus also acknowledges support from the Japan Society for
the Promotion of Science and FWO-Vl. A. D. Hern\'andez also acknowledges
support from the Centro Latinoamericano de F\'{\i}sica and Fundaci\'on
Antorchas.

\end{document}